# Metrics for Bengali Text Entry Research

**Sayan Sarcar**

Center for Human Computer Interaction

Kochi University of Technology

Kami, Kochi, Japan

mailtosayan@gmail.com

**Ahmed Sabbir Arif**

Synaesthetic Media Laboratory

Ryerson University

Toronto, Ontario, Canada

asarif@ryerson.ca

**Ali Mazalek**

Synaesthetic Media Laboratory

Ryerson University

Toronto, Ontario, Canada

mazalek@ryerson.ca

## Abstract

With the intention of bringing uniformity to Bengali text entry research, here we present a new approach for calculating the most popular English text entry evaluation metrics for Bengali. To demonstrate our approach, we conducted a user study where we evaluated four popular Bengali text entry techniques.

## Author Keywords

Bangla/Bengali text entry; performance metrics.

## Introduction

Bengali text entry on handheld devices is becoming increasingly popular among native speakers [8], as it allows them to communicate with each other without the assistance of a second language. There are numerous Bengali text entry techniques available for handheld devices. There are virtual keyboards that either map different keyboard layouts onto QWERTY (INSCRIPT keyboards, i.e. Bangla Jatiyo, Probhat, and Bijoy [2]) or develop novel keyboard layouts (i.e. Swarachakra [6], iLiPi-B [9], and Bhattacharya [4]). There are phonetic keyboards that use phonetic mapping schemes to support Bengali text entry with Roman characters (i.e. Akkhor, Avro, Google Bengali Transliteration [5], Microsoft Bengali Transliteration, and Onkur). There are techniques that utilize the standard 12-key mobile keypad (i.e. Panini [7] and Nokia Bengali). There are also several techniques that attempt to assign dedicated keys for all Bengali characters.

In the Bengali language there are approximately eleven vowels, thirty-six consonants, ten inflexions, four signs,

and ten numeric characters. Besides, similar to other Indic languages such as Hindi, Bengali supports combinations between consonants (conjunct) and consonants and dependent vowels (glyph). Although, visually the word কাণ্ড appears to be consisted of two characters কা and ণ্ড, in reality the former is a glyph (ক+া) and the latter is a conjunct (ণ+্+ড). Therefore, to enter this word with a character-based technique, one would need to input the following five characters in sequence ক+া+ণ+্+ড. Similarly, it would require five backspaces to delete it. As Bengali has more characters than English, INSCRIPT techniques assign multiple characters to a single key, and then require the use of special keys (i.e. Shift) or key combinations (i.e. Shift + Ctrl) to disambiguate between them. Yet there is no commonly accepted standard for this. Similarly, Phonetic keyboards use different phonetic mapping schemes and character sequence to create conjuncts and glyphs. For example, based on the input sequence used by the scheme, প+্+র could yield either প্র or র্প. Besides, some keyboards provide dedicated keys for the most frequent conjuncts (i.e. ক্ষ, জ্ঞ, etc.). Therefore, the process of Bengali text entry, editing, and error correction is fundamentally different with different techniques.

Researchers usually use the most popular performance metrics for English text entry research, such as words per minute (WPM), keystrokes per character (KSPC), and minimum string distance error rate (MSDER), to evaluate Bengali text entry techniques. As these metrics were not designed for the Bengali language and are not equipped to address the diversity of Bengali text entry techniques, researchers and practitioners are forced to use different conventions to calculate them. This makes it difficult to evaluate and compare different Bengali text entry techniques, as these metrics usually yield different results for the same scenario based on how characters were defined and the sequence in which the conjuncts and the glyphs were generated. For instance, based on whether a conjunct was counted as a single or multiple character(s), these metrics will yield different results for the same input/output. To address this, here we propose a new method for calculating these metrics for Bengali.

**Proposed Methodology**

Here we discuss the new approach for calculating the most common text entry performance metrics, namely WPM, KSPC, ER, and MSDER, for Bengali text entry. We use the following notations [1, 10].

- *Input Stream (IS)* is the text containing all keystrokes produced by the user while entering the presented text. For example, if the user performs the keystrokes b+a+i with a transliteration system to input the Bengali word বই, then the input stream will also contain b+a+i. |IS| is the length of the input stream.
- *Output Stream (OS)* is the text that contains all "constituent" characters in a phrase. In other words, it considers the outputted characters instead of the exact sequence of keystrokes performed. Thus, for the above example, the output stream will contain ব+ই. Also, as it disjoins all conjuncts and glyphs into basic characters, even if the user inputs the conjunct ক্ষ (ক+্+ষ) directly, such as by pressing a dedicated key for it, the output stream will contain ক+্+ষ.
- *Presented Text (P)* is what participants had to enter, $OS_P$ the presented text's output stream, and $|OS_P|$ is the length of $OS_P$.
- *Transcribed Text (T)* is the final text entered by the participant, $OS_T$ is the transcribed text's output stream, and $|OS_T|$ is the length of $OS_T$.
- *Incorrect Not Fixed ($INF_{OS}$)* is the total number of uncorrected characters in the transcribed text's OS.

**Sample Calculation**

Here we calculate the proposed metrics for an example scenario where the examined technique has a dedicated key for the conjunct ক্ষ (ক+্+ষ)

Presented Text (P):
ক্ষণিকের অতিথি

Output Stream (OS$_P$):
ক+্+ষ+ণ+ি+ক+ে+র+ +অ+ত+ি+থ+ি

|OS$_P$| = 14

Transcribed Text (T):
ক্ষণিকের অতথি

Output Stream (OS$_T$):
ক+্+ষ+ণ+ি+ক+ে+র+ +অ+ত+থ+ি

|OS$_T$| = 13

Input Stream (IS):
ক্ষ+Shift+ণ+ি+ক+ে+র+ +অ+ত+থ+ি

|IS| = 12

---

INF$_{OS}$ = 1
MSD$_{OS}$ = 1
Let us assume, S = 20 seconds

---

- $WPM_{bn} = \frac{13-1}{20} \times 60 \times \frac{1}{5.11} = 7.05$
- $KSPC_{bn} = \frac{12}{13} = 0.92$
- $ER_{bn} = \frac{1}{13} \times 100 = 7.69\%$
- $MSDER_{bn} = \frac{1}{14} \times 100 = 7.14\%$

---

- *Minimum String Distance (MSD$_{OS}$)* is the minimum number of operations needed to transform the transcribed text's output stream to the presented text's output stream, where the operations are insertion, deletion, or substitution of a basic character. If a technique allows the insertion, deletion, or substitution of a conjunct or a glyph, then the number of operations required is one over the number of characters in the conjunct or glyph. For instance, if it allows the user to delete a conjunct consisted of four basic characters using one backspace, then the number of operations required is ¼ = 0.25.

*Words per Minute for Bengali (WPM$_{bn}$)*
WPM$_{BN}$ measures the time it takes to produce a certain number of Bengali words.

$$WPM_{bn} = \frac{|OS_T|-1}{S} \times 60 \times \frac{1}{5.11} \quad (1)$$

In Equation 1, S is time in seconds measured from the first keystroke to the last, including backspaces and other edit and modifier keys. The constant 60 is the number of seconds per minute, and 1/5.11 accounts for the average length of a word in characters including spaces, numbers, and other printable characters in Bengali [3]. As S is measured from the entry of the very first character to the last, the entry of the first character is never timed, which is the motivation for the –1 in the numerator.

*Keystrokes per Character for Bengali (KSPC$_{bn}$)*
KSPC$_{bn}$ is the ratio of the length of the input stream to the length of the transcribed text's output stream.

$$KSPC_{bn} = \frac{|IS|}{|OS_T|} \quad (2)$$

*Error Rate for Bengali (ER$_{bn}$, %)*
ER$_{bn}$ is the ratio of the total number of incorrect characters in the transcribed text's output stream to the length of the transcribed text's output stream.

$$ER_{bn} = \frac{INF_{OS}}{|OS_T|} \times 100 \quad (3)$$

*MSD Error Rate for Bengali (MSDER$_{bn}$, %)*
MSDER$_{bn}$ calculates the smallest number of operations to transform the transcribed text's output stream to match the presented text's output stream, and then calculates the ratio of that number to the larger of the length of the presented and transcribed text's output stream.

$$MSDER_{bn} = \frac{MSD_{OS}}{Max(|OS_P|,|OS_T|)} \times 100 \quad (4)$$

*Other Performance Metrics*
We could also calculate other popular text entry performance metrics, such as Erroneous Keystrokes [1] and Total Error Rate [10], using this approach.

## Current Methods

Researchers use various methods to calculate these metrics. Yet, as none of them deal with conjuncts and glyphs in a meaningful way and do not differentiate between the input and the output, they usually fail to show an accurate picture. One popular approach is to consider conjuncts and/or glyphs as individual characters instead of combinations. As a result, if it takes about the same time to enter two different phrases with two different techniques, one containing substantially more conjuncts than the other, this approach incorrectly assumes that both as equally fast—although, the conjunct-heavy phrase required substantially more character entry than the other. Similarly, this method assumes that the difference between বৃক্ষ and ব্রক্ষ is one character, although with most techniques replacing বৃ (ব+ৃ) with ব্র (ব+্+র) require correcting two unwanted characters (্+র).

## A User Study

We conducted a user study to evaluate four popular Bengali text entry techniques: Swarachakra, Panini,

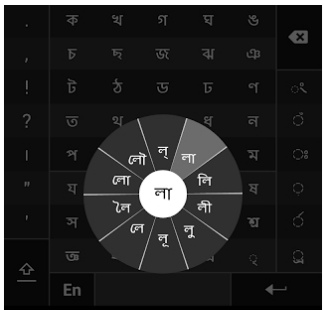

(a) Swarachakra

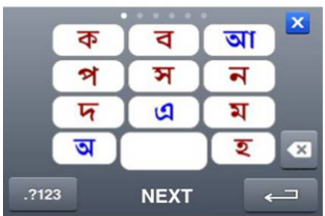

(b) Panini

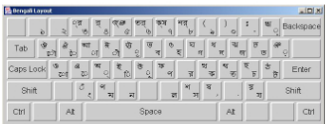

(c) Indic/InScript

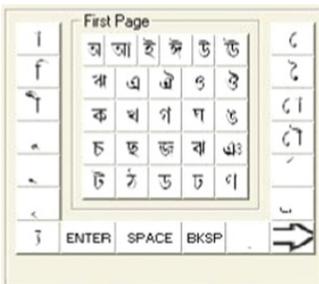

(d) Bhattacharya

**Figure 1.** Screenshots of the evaluated techniques.

InScript, and Bhattacharya (see Figure 1), with our methodology. The study involved inputting thirty short Bengali phrases (randomly collected from different books) with each technique on a touchscreen-based smartphone (Samsung Galaxy S3). Fifteen subjects voluntarily participated in the study. Their age ranged from 22 to 31 years, average 26 (SD = 4). Six of them were female and they all had prior experiences with touchscreen-based Bengali text entry. They were asked to enter the phrases as fast and accurate as possible. In summary, the design was: 15 participants × 4 techniques (counterbalanced) × 30 short Bengali phrases = 1,800 phrases in total.

| Techniques | $WPM_{bn}$ | $KSPC_{bn}$ | $ER_{bn}$ | $MSDER_{bn}$ |
|---|---|---|---|---|
| Swarachakra | 10.07 | 6.23 | 10.05% | 1.22% |
| Panini | 15.02 | 7.86 | 14.99% | 1.35% |
| Indic/INSCRIPT | 14.10 | 5.08 | 14.08% | 1.33% |
| Bhattacharya | 7.69 | 6.25 | 7.68% | 1.67% |

**Table 1.** Text entry performance for four popular Bengali text entry techniques.

Results (see Table 1) show that Panini and Indic were substantially faster than the other techniques, but were also unreliable. However, text entry with Indic required about 35% fewer keystrokes than Panini. Swarachakra yielded a mediocre performance, both in terms of speed and accuracy, while Bhattacharya's performance was the poorest. Our approach allowed us to measure and compare these techniques' performance, although they are substantially different from one another.

## Conclusion
With the intention of bringing uniformity in Bengali text entry research, we presented and demonstrated a new approach for calculating the most popular English text entry evaluation metrics for Bengali.

## Future Work
As most Indic languages (i.e. Hindi, Punjabi, Marathi, Gujarati, Bhojpuri, Oriya, Sindhi, Sinhala, Nepali, Assamese, etc.) use a similar writing system, we speculate that this approach could be used with these languages as well. We would like to investigate this in the future.

## References
[1] Arif, A. S. and Stuerzlinger, W. Analysis of text entry performance metrics. *TIC-STH '09*. IEEE, 100-105.

[2] Bengali Input Methods. http://en.wikipedia.org/wiki/Bengali_input_methods

[3] Bharati, A., Rao, P., Sangal, R. and Bendre, S. M. Basic statistical analysis of corpus and cross comparison among corpora. Technical Report ('00), IIIT.

[4] Bhattacharya S., and Laha S. Bengali text input interface design for mobile devices. *UAIS 12*, 4 ('13), 441-451.

[5] Google InScript and Phonetic. http://www.google.com/intl/bn/inputtools/try

[6] Joshi, A., Dalvi, G., Joshi, M., Rashinkar, P., and Sarangdhar, A. Design and evaluation of Devanagari virtual keyboards for touch screen mobile phones. *MobileHCI '11*. ACM, 323–332.

[7] Panini Keypad. http://paninikeypad.com/index1.php

[8] Prince, M. E. H., Hossain, G., Dewan, A. A., and Debnath, P. An audible Bangla text-entry method in mobile phones with intelligent keypad. *ICCIT '09*. IEEE, 279-284.

[9] Samanta, D., Sarcar, S., and Ghosh, S. An approach to design virtual keyboards for text composition in Indian languages. *International Journal of Human Computer Interaction 29*, 8 (2013), 516-540.

[10] Soukoreff, R. W. and MacKenzie, I. S. Metrics for text entry research: an evaluation of MSD and KSPC, and a new unified error metric. *CHI '03*. ACM, 113-120.